\documentclass[12 pt]{article}
\usepackage{times,bm}
\usepackage[affil-it,]{authblk}
\usepackage[pagebackref=false,colorlinks=true,linkcolor=blue,citecolor=blue,urlcolor=blue]{hyperref}
\usepackage{geometry}
\usepackage{graphicx}
\usepackage{cite}
\usepackage{amsmath}
\usepackage{amsfonts}
\usepackage{soul}
\usepackage{amssymb}
\usepackage{subfigure}
\usepackage{color}
\usepackage{float}
\usepackage{multirow,multicol}
\usepackage{tabulary}
\usepackage[flushleft]{threeparttable}
\usepackage{pgfplots,subfigure}
\usepackage{xcolor}
\numberwithin{equation}{section}
\usepackage{tikz}
\usepackage{placeins}
\usepackage{caption}
\usepackage[normalem]{ulem}
\usepackage{hyperref}
\usepackage{nameref}
\usepackage{comment}
\usepackage{pgfplots}
\pgfplotsset{compat=1.7}

\usepgflibrary{arrows}
\usetikzlibrary{shapes.callouts}
\tikzset{
	level/.style   = { thick, },
	connect/.style = { dotted, red   },
	notice/.style  = { draw, rectangle callout, callout relative pointer={#1} },
	label/.style   = { text width=2cm }
}

\usepackage{letltxmacro}
\makeatletter
\let\oldr@@t\r@@t
\def\r@@t#1#2{%
	\setbox0=\hbox{$\oldr@@t#1{#2\,}$}\dimen0=\ht0
	\advance\dimen0-0.2\ht0
	\setbox2=\hbox{\vrule height\ht0 depth -\dimen0}%
	{\box0\lower0.4pt\box2}}
\LetLtxMacro{\oldsqrt}{\sqrt}
\renewcommand*{\sqrt}[2][\ ]{\oldsqrt[#1]{#2}}
\makeatother

\begin{document}
	\newcommand{{\ri}}{{\rm{i}}}
	\newcommand{{\Psibar}}{{\bar{\Psi}}}
	\newcommand{{\red}}{\color{red}}
	\newcommand{{\blue}}{\color{blue}}
	\newcommand{{\green}}{\color{green}}
	\newcommand{\rev}[1]{\textbf{\textcolor{red}{#1}}}
	
	\title{Scattering effects of bumblebee gravity in metric–affine
formalism}

	\author{\large  
		\textit {N. Heidari}$^{\ 1}$ \footnote{E-mail: heidari.n@gmail.com (Corresponding author)},
		\textit {Caio F. B. Macedo}$^{\ 2}$\footnote{E-mail: caiomacedo@ufpa.br},
		\textit{A. A. Ara\'{u}jo Filho}$^{\ 3}$
		\footnote{E-mail: dilto@fisica.ufc.br}	and 
		\textit {H.Hassanabadi}$^{\ 1,4 }$\footnote{E-mail: hha1349@gmail.com},
		\\
		\small \textit {$^{\ 1}$Faculty of Physics, Shahrood University of Technology, Shahrood, Iran.}\\
		\small \textit {$^{\ 2}$Faculdade de Física, Campus Salinópolis, Universidade Federal do Pará, 68721 -- 000, Salinópolis, Pará, Brazil.}\\

        \small \textit {$^{\ 3}$Departamento de Física, Universidade Federal da Paraíba, Caixa Postal 5008, 58051--970, João Pessoa, Paraíba,  Brazil.}\\
		
		\small\textit {$^{\ 4}$ Department of Physics, University of Hradec Kr$\acute{a}$lov$\acute{e}$, Rokitansk$\acute{e}$ho 62, 500 03 Hradec Kr$\acute{a}$lov$\acute{e}$, Czechia.}
	}
	
	\date{}
	\maketitle

	\begin{abstract}
		In this work, we explore a Schwarzschild-like black hole within the framework of metric--affine bumblebee gravity. First, we investigate the behavior of the Kretschmann scalar and singularities in this modified gravity approach. Next, we introduce a newly defined time coordinate related to a stationary asymptotically flat spacetime. We also analyze the scattering effects and numerically calculate and comprehensively examine the partial and total absorption cross sections. At the high--frequency approximation, we find that the absorption cross section tends to the geodesic capture cross section.  The continued fraction method is applied to investigate the quasinormal modes, and we explore the deviations of both the real and imaginary terms of the quasinormal modes from the Schwarzschild case in detail. We verify the relation between the shadow radius and the real part of the quasinormal frequencies at the eikonal limit within this modified gravity framework. Finally, we examine the energy emission rate.

	\end{abstract}
	
	\begin{small}
	\end{small}
	
	\FloatBarrier
	
	
	\section{Introduction}

Lorentz symmetry, based on the principles of Special Relativity, is fundamental for ensuring observational equivalence. It ensures that physical laws remain consistent for all observers in inertial frames. This symmetry, involving both rotational and boost aspects, is essential in general relativity and the standard model of particle physics. In curved spacetimes, Lorentz symmetry appears locally, reflecting the Lorentzian nature of the background. However, when inertial frame conditions are not met, subtle dependencies on direction or velocity arise, altering the dynamics of particles and waves \cite{STR1,STR2,STR3,STR4,STR5,STR6,STR7}.

Processes of symmetry breaking consistently reveal intriguing consequences and often signal new physical phenomena. Specifically, Lorentz symmetry breaking (LSB) leads to a variety of distinctive features \cite{liberati2013,tasson2014,hees2016}, providing valuable insights into quantum gravity \cite{rovelli2004}. Numerous theoretical models operate on the assumption of Lorentz invariance violation, including closed string theories \cite{New1,New2,New3,New4,New5}, loop quantum gravity \cite{New6,New7}, noncommutative spacetimes \cite{New8,New9}, non--local gravity models \cite{Modesto:2011kw, Nascimento:2021bzb}, spacetime foam models \cite{New10,New11}, and (chiral) field theories on spacetimes with nontrivial topologies \cite{New12,New13,New14,New15}, along with Hořava-Lifshitz gravity \cite{New16} and cosmology \cite{sv1,sv2}.

Exploring thermal radiation within the context of Lorentz symmetry breaking (LSB) provides valuable insights into the characteristics of the primordial Universe. This perspective is based on the observation that, during this early stage, the size of the Universe matches the characteristic scales of Lorentz violation \cite{kostelecky2011data}. The investigation into the thermal properties related to LSB was initially proposed in \cite{colladay2004statistical}. Since then, numerous studies have examined various scenarios, including linearized gravity \cite{aa2021lorentz}, Pospelov and Myers--Pospelov electrodynamics \cite{araujo2021thermodynamic,anacleto2018lorentz}, CPT--even and CPT--odd LV terms \cite{casana2008lorentz,casana2009finite,araujo2021higher,aguirre2021lorentz}, higher-dimensional operators \cite{Mariz:2011ed, reis2021thermal}, bouncing universe models \cite{petrov2021bouncing2}, \textit{rainbow} gravity \cite{furtado2023thermal}, and Einstein--aether theory \cite{aaa2021thermodynamics}. These studies collectively enhance our understanding of the thermal implications of LSB within various theoretical frameworks. Moreover, integrating Lorentz symmetry breaking into gravitational frameworks presents unique challenges distinct from incorporating Lorentz violation (LV) extensions in non-gravitational field theories. In flat spacetimes, additive LV terms, such as the Carroll--Field--Jackiw \cite{CFJ} and aether terms \cite{aether, Gomes:2009ch}, can be introduced seamlessly. For a comprehensive approach encompassing all possible LV minimal couplings, refer to \cite{colladay1998lorentz}.

The Standard Model Extension (SME), encompassing its gravitational sector, stands as a concise and inclusive framework that comprehensively addresses all potential coefficients for Lorentz/CPT violation \cite{5,kostelecky2021backgrounds,bluhm2003probing,bluhm2005spontaneous,bluhm2008spontaneous,bluhm2021gravity,bluhm2023spontaneous,kostelecky2011matter}. Specifically, within its gravitational realm, the SME operates on a Riemann--Cartan manifold, dynamically considering torsion as a geometric quantity alongside the metric. Although there is room to introduce non--Riemannian terms in the gravity SME sector, current research has predominantly emphasized the metric approach to gravity, where the metric serves as the singular dynamical geometric field.

In this context, significant efforts have focused on deriving exact solutions for various models that incorporate Lorentz symmetry breaking (LSB) within curved spacetimes. Notable examples include studies on bumblebee gravity using the metric approach \cite{6,7,8,9,10,11,12,13,14,Maluf:2021lwh, KumarJha:2020ivj}, the Einstein--aether model \cite{15}, parity--violating models \cite{16,17,18,19,20,Rao:2023doc}, and Chern--Simons modified gravity \cite{21,22,23}. Experimental tests have also been conducted to detect signals of LSB in the weak field regime of gravitational fields, with Solar System experiments being particularly notable \cite{24,25,26}. The recent detection of gravitational waves by the LIGO/VIRGO collaboration \cite{LIGOScientific:2016aoc}, enabled by advancements in technology, has opened a new window for investigating the strong field regime of gravity. This development offers a powerful tool for exploring the complex properties of compact objects, such as black holes, whose shadows have already been captured in images \cite{EventHorizonTelescope:2019dse, EventHorizonTelescope:2022wkp}.

While many studies explore modified theories of gravity using the traditional metric approach, there is increasing interest in more versatile geometrical frameworks. Notably, examining theories of gravity within a Riemann--Cartan background offers specific advantages, such as the induction of gravitational topological terms \cite{Nascimento:2021vou}. Another intriguing non--Riemannian geometry is Finsler geometry \cite{Bao}, which has been extensively linked to Lorentz symmetry breaking (LSB) in various studies \cite{Foster, KosE, Sch1, CollM, Sch2}.

The metric--affine (Palatini) formalism offers a notable generalization of the metric approach by treating the metric and connection as independent dynamical quantities (for detailed exploration and findings within the Palatini approach, see \cite{Ghil1,Ghil2}). Despite advancements in this framework, Lorentz symmetry breaking has remained relatively unexplored. However, recent efforts have begun to address this gap, particularly in the context of bumblebee gravity scenarios \cite{Paulo2, Paulo3, Paulo4}. These studies have led to the derivation of field equations, generic solutions, and an examination of stability conditions and associated dispersion relations for various matter sources within the weak field and post--Newtonian limits.

On the quantum front, the divergent part of one--loop corrections to the spinor effective action has been computed using two distinct methods: the diagrammatic method in the weak gravity regime and the Barvinsky-Vilkovisky technique on a broader scale. A significant breakthrough has been the discovery of exact Schwarzschild--like \cite{Filho:2022yrk} and Kerr--like \cite{araujo2024exploring} solutions, allowing for estimations of the Lorentz violation parameter from classical gravitational tests. Additionally, comprehensive studies have recently been conducted on the shadow and quasinormal modes of these black holes \cite{Lambiase:2023zeo, Jha:2023vhn, araujo2024gravitational, nascimento2024gravitational}.

Although many studies concerning the metric from Ref. \cite{Filho:2022yrk} have been conducted, several aspects remain unexplored, leaving gaps in the literature. This research addresses these gaps by investigating the behavior of the Kretschmann scalar and singularities within this modified gravity framework. A newly defined time coordinate for a stationary asymptotically flat spacetime is introduced. Scattering effects are analyzed, and both the partial and total absorption cross sections are numerically calculated and thoroughly examined. In the high-frequency approximation, the absorption cross section approaches the geodesic capture cross section. The continued fraction method is used to investigate the quasinormal modes, providing detailed deviations of both the real and imaginary components from the Schwarzschild case. The correlation between the shadow radius and the real part of the quasinormal frequencies at the eikonal limit is verified. Finally, the energy emission rate is also examined.

This work is organized as follows: In Sec. \ref{pan}, the Kretschmann scalar is investigated. Sec. \ref{part} explores the partial wave function and considers the impact of the bumblebee parameter on the effective potential. Sec. \ref{abs} analyzes the partial and total absorption cross sections through numerical calculations. The greybody bound is calculated in Sec. \ref{Tbound}. The effect of bumblebee spacetime on the quasinormal modes is examined using Leaver’s continued fraction method in Sec. \ref{QNMs}, where the correspondence between quasinormal modes and the shadow radius is also discussed. The energy emission rate is explored in Sec. \ref{EMR}. Finally, concluding remarks are provided in Sec. \ref{conclusion}. 


\section{General panorama}\label{pan}

The metric derived in \cite{Filho:2022yrk} characterizes a non-rotating black hole with incorporated Lorentz symmetry breaking effects. The form of the metric was demonstrated to exhibit 
\begin{equation}
    \mathrm{d}s^2_{(g)}=-\frac{\left(1-\frac{2M}{r}\right)}{\sqrt{\left(1+\frac{3X}{4}\right)\left(1-\frac{X}{4}\right)}}\mathrm{d}t^2+\frac{\mathrm{d}r^2}{\left(1-\frac{2M}{r}\right)}\sqrt{\frac{\left(1+\frac{3X}{4}\right)}{\left(1-\frac{X}{4}\right)^3}}+r^{2}\left(\mathrm{d}\theta^2 +\sin^{2}{\theta}\mathrm{d}\phi^2\right),
    \label{metric3}
\end{equation}
where \(X = \xi b^2\) serves as the Lorentz violation coefficient shorthand, previous studies on this black hole have addressed various aspects, such as quasinormal modes, \textit{Hawking} radiation \cite{Jha:2023vhn} and accretion disk\cite{lambiase2023probing}. 
It is important to mention that, within the asymptotic limit, the time components of the metric read: $\lim\limits_{r \to \infty} g_{tt} = \frac{1}{\sqrt{\left(1+\frac{3X}{4}\right)\left(1-\frac{X}{4}\right)}}$. Given that \(X\) is a constant, we can perform a coordinate rescaling of \(t\) as follows: \( \left[ \left( 1 + \frac{3X}{4}    \right) \left(  1 - \frac{X}{4}  \right) \right]^{-1/4} t  \to   \tilde{t}\), where $\tilde{t}$ will be the time related to a stationary asymptotic observer.
For the sake of convenience, let us define $f(r) \equiv 1 - \frac{2M}{r}$, $\alpha \equiv \frac{1}{\sqrt{\left(1+\frac{3X}{4}\right)\left(1-\frac{X}{4}\right)}}$, $\beta \equiv \sqrt{\frac{\left(1+\frac{3X}{4}\right)}{\left(1-\frac{X}{4}\right)^{3}}}$. In other words, we work with the following line element
\begin{equation}\label{dsbar}
    \mathrm{\rm{d}}\tilde{s}^2=-f(r)\mathrm{\rm{d}}\tilde{t}^2+\frac{\beta}{f(r)}  \mathrm{\rm{d}}r^2+r^{2}\left(\mathrm{\rm{d}}\theta^2 +\sin^{2}{\theta}\mathrm{\rm{d}}\phi^2\right).
\end{equation}

\subsection{Kretschmann scalar}\label{kret}

\begin{figure}[h]
	\centering
	\includegraphics[width=80mm]{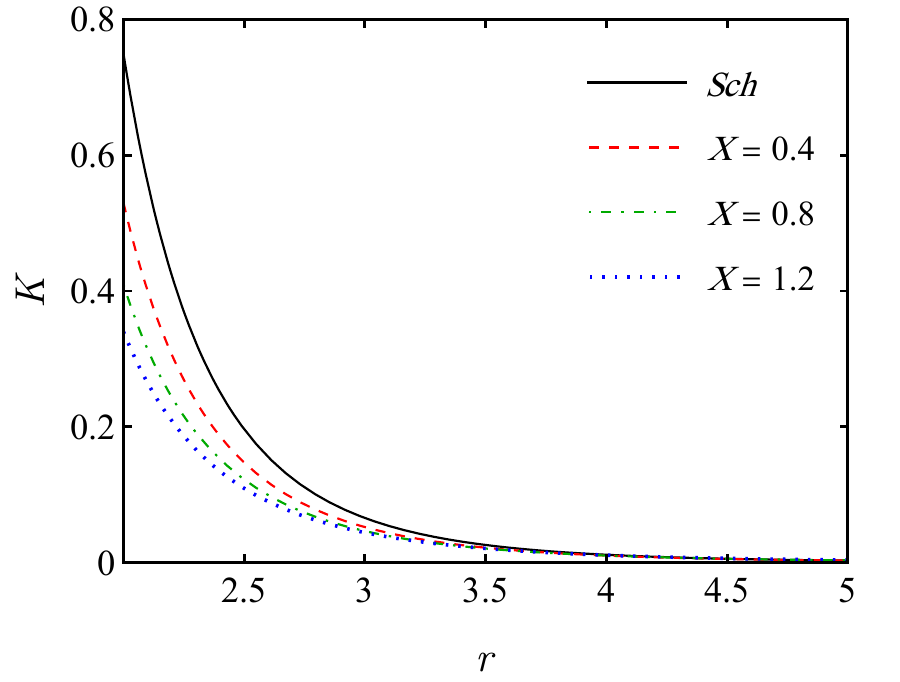} 
	\caption{Illustration of the Kretschmann scalar against the radius $r$. The plot is obtained for $M=1$}
	\label{fig:kresh}
\end{figure}

In order to analyze the behavior of the spacetime previously introduced, the Kretschmann scalar is discussed. In this manner, it reads
\begin{equation}
	K = {R_{\mu \nu \gamma \lambda }}{R^{\mu \nu \gamma \lambda }}.
\end{equation}
Here, $R_{\mu \nu \gamma \lambda }$ is the Reinmann tensor components, which follows
\begin{equation}\label{kretsch}
	K=\frac{4 \left(12 M^2+4 (\beta -1) M r+(\beta -1)^2 r^2\right)}{\beta ^2 r^6}.
\end{equation}
By considering $X=0$ $(\beta=1)$, the Kretschmann scalar approaches to $\frac{48M^2}{r^6}$, representing expected expression for a Schwarzschild black hole, as one can naturally verify. Such a scalar versus radius $r$ is plotted for some values of parameter $X$ in Fig. \ref{fig:kresh}.
The plots depict that the asymptotic behavior of the metric is the same as the Schwarzschild spacetime. To examine the existence of a possible singularity, we set Eq. \eqref{kretsch} to be zero so that $r =  - \frac{{2M}}{{(\beta  - 1)}}(1 \pm i\sqrt 2 ) = - 2M(1 \pm i\sqrt 2 ){(\sqrt {\left( {1 + \frac{{3X}}{4}} \right){{\left( {1 - \frac{X}{4}} \right)}^{ - 3}}}  - 1)^{ - 1}}$. This means that the Kretschmann scalar will be zero only on a complex radius (non--physical) exists. Therefore, $K$ possess a singularity in $r=0$ only, analogous to the Schwarzschild black hole.

\section{partial wave equation}\label{part}

In this section, the partial wave equation will be explored to find the scalar filed quasinormal modes.

Now, let us consider the Klein--Gordon equation in curved spacetime introduced in Eq. \eqref{dsbar} as
\begin{equation}\label{klein}
	\frac{1}{{\sqrt { - g} }}{\partial _\mu }(\sqrt { - g} {g^{\mu \nu }}{\partial _\nu }\Psi ) = 0,
\end{equation}
and applying the separation of variables, we have
\begin{equation}\label{sai}
	{\Psi _{\omega lm}}(\mathbf{r},t) = \frac{{{\psi_{\omega l}}(r)}}{r}{Y_{lm}}(\theta ,\varphi ){e^{ - i\omega \tilde{t}}}.
\end{equation}
In a general form of spherically symmetric spacetime as $\mathrm{d}{s^2} =  - g_{tt}\mathrm{d}{\tilde{t}^2} + {g_{rr}}\mathrm{d}{r^2} + r^2 \mathrm{d}{\Omega ^2} $, we can define the tortoise coordinate 
\begin{equation}
	{\rm{d}}r^* = \frac{{\rm{d}}r}{\sqrt {|g_{tt}|g_{rr}^{ - 1}} } =\sqrt{\beta}\frac{\rm{d}r}{f(r)},
\end{equation}
so that the Klein--Gordon equation can be reduced to a Schrödinger--like wave function
\begin{equation}\label{waves}
	\left[\frac{{{\rm{d}^2}}}{{\rm{d}{{r^*}^2}}} + ({\omega ^2} - {V_{eff}})\right]\psi_{\omega l} (r) = 0,
\end{equation}
where the $V_{eff}$ is the effective potential and its explicit form is give below
\begin{equation}
\label{Veff}
\begin{split}
V_{eff}& = |{g_{tt}}|\left[\frac{{l(l + 1)}}{{{r^2}}} + \frac{1}{{r\sqrt {|{g_{tt}}|{g_{rr}}} }}\frac{\mathrm{d}}{{\mathrm{d}r}}\sqrt {|{g_{tt}}|g_{rr}^{ - 1}}\right]\\ 
&=f(r)\left[ \frac{l\left( {1 + l} \right)}{r^2}  + \sqrt{\frac{\left(1-\frac{X}{4}\right)^3}{\left(1+\frac{3X}{4}\right)}} \frac{1}{r} \frac{\mathrm{d}f(r)}{\mathrm{d}r}  \right].
\end{split}
\end{equation}

\begin{figure}[h]
	\centering
	\includegraphics[width=80mm]{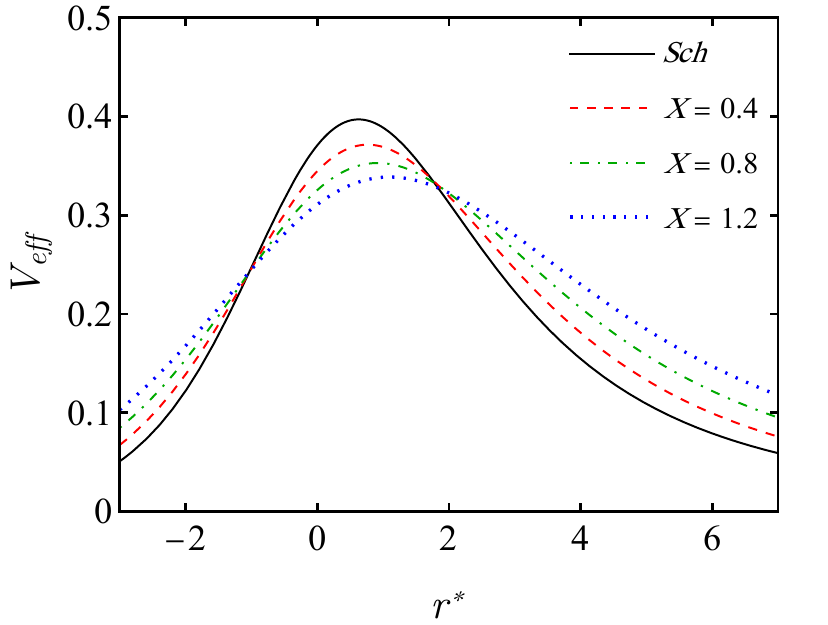} 
	\caption{The effective potential is shown for $M = 0.5$, $l=1$ and different values of $X$.}
	\label{fig:Veff}
\end{figure}

For the sake of providing a better comprehension to the reader, $V_{eff}$ against the tortoise coordinate is represented in Fig. \ref{fig:Veff} for some values of parameter $X$, i.e., including the Schwarzschild case ($X=0$). Notice that parameter $X$ impacts the height of effective potential; for higher values of bumblebee parameter $X$, the barrier of $V_{eff}$ decreases. 

\section{Absorption cross section}\label{abs}

As shown in Fig. \ref{fig:Veff}, the scalar field's effective potential is localized, vanishing at both limits of $r^{*}$ which correspond to the horizon and infinity. Considering its characteristics, the asymptotic behavior of the radial wavefunction is expected to be purely
incoming at the event horizon and composition of ingoing and outgoing waves at spatial infinity and can be described as the following boundary conditions \cite{macedo2015scattering,macedo2016absorption} 
\begin{equation}\label{bound}
	{\psi _{\omega l}} \approx 
	\begin{cases}
	\mathcal{T}_{\omega l} R_{{1}} , &\text{for} \ {r^*} \to +r_h ~ (r \to -\infty )\\
	R_{{2}}+\mathcal{R}_{\omega l} R^*_{{2}},&\text{for} \ {r^*} \to +\infty ~ (r \to \infty )\\
	\end{cases}
\end{equation}
where $|\mathcal{R}_l|^2$ and $|\mathcal{T}_l|^2$  are the reflection and transmission coefficients, respectively.  Due to the flux conservation, this relation $|\mathcal{R}_l|^2+|\mathcal{T}_l|^2=1$ is satisfied.
Moreover, $R_{{1}}$ and $R_{{2}}$ can be written as \cite{macedo2013absorption}  
\begin{align}\label{Rroman1}
	{R_{{1}}}& = {e^{ - i\omega r*}}\sum\limits_{j = 0}^N {{{(r - {r_h})}^j}A_{{r_h}}^{(j)}}, \\ \label{Rroman2}
	{R_{{2}}}&= {e^{ -i\omega r*}}\sum\limits_{j = 0}^N {\frac{{A_\infty ^{(j)}}}{{{r^j}}}} .
\end{align}
To find the coefficients $A_\infty ^{(j)}$ and $A_{{r_h}}^{(j)}$, it is necessary to require that the functions ${R_{{1}}}$ and ${R_{{2}}}$ satisfy the differential equation in Eq. \eqref{waves} when far from the black hole and close to the event horizon, respectively.
The phase shift $\delta_{\omega l}$ is defined as \cite{dolan2009scattering}
\begin{equation}\label{phase}
	{e^{2i{\delta _{\omega l}}}} = {( - 1)^{l + 1}}\mathcal{R}_{\omega l}.
\end{equation}	
To obtain the phase shifts for computing the absorption cross--section, we followed the numerical method discussed in Ref. \cite{crispino2009scattering} to solve
the radial equation \eqref{waves}. We have considered Eq. \eqref{Rroman1} and Eq. \eqref{Rroman2} and evaluated the series at 10th order $(N = 10)$. Started from near the horizon at $\frac{r}{{{r_h}}} - 1 = {10^{ - 3}}$ into the asymptotically flat region at $r \sim 200{r_h}$, we matched the numerical solution onto the boundary conditions in Eq. \eqref{bound}.

On the other hand, the total absorption cross section can be written as $\sigma _{abs} = \sum\limits_{l = 0}^\infty  {\sigma _{abs}^l}$, where $\sigma_l$ is the partial absorption cross, being described by \cite{crispino2009scattering}
\begin{equation}\label{partial1}
		\sigma _{abs}^l = \frac{\pi }{{{\omega ^2}}}(2l + 1)(1 - {\left| {{e^{2i{\delta _{\omega l}}}}} \right|^2}).
	\end{equation}
By applying Eq. \eqref{phase} and the relation of flux conservation, we arrive at ${\left| {{e^{2i{\delta _{\omega l}}}}} \right|^2}=1-|\mathcal{R}_{\omega l}|^2={\left| {{T_{\omega l}}} \right|^2}$ \cite{crispino2013greybody}; moreover, due to the asymptotic form of the metric, we have that the plane wave is $\propto e^{-i\omega \sqrt{\beta}r}$. This makes the absorption cross section have an extra factor $\beta$ and can be written as
\begin{equation}\label{partial2}
	{\sigma _{abs}} = \frac{\pi }{{{\beta\omega ^2}}}\sum\limits_{l = 0}^\infty  {(2l + 1)} {\left| {{T_{\omega l}}} \right|^2},
\end{equation}
where ${\left| {{T_{\omega l}}} \right|^2}$ is transmission coefficient called greybody factor.
The partial absorption cross sections for different values of $X$ can be calculated by applying the numerical transmission coefficients into Eq. \eqref{partial1}. 

\begin{figure}[h]
	\centering
	\includegraphics[width=80mm]{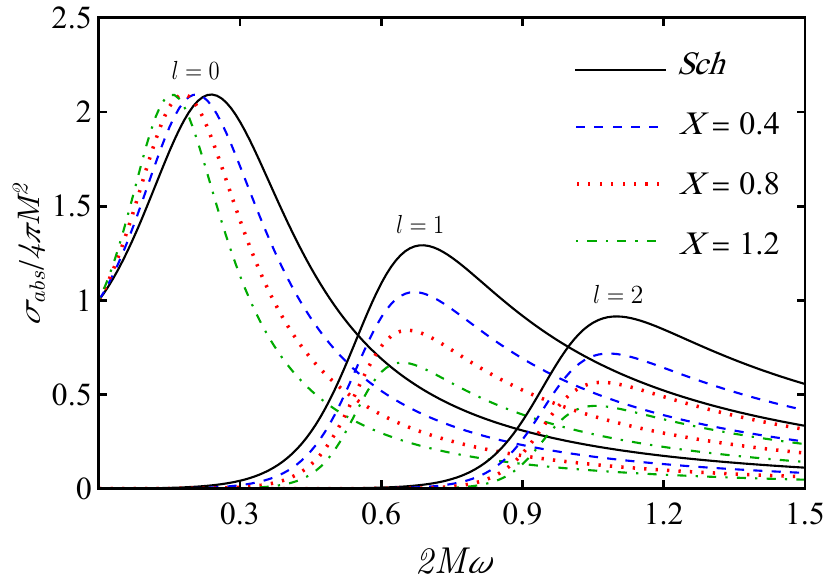} 
	\caption{Partial absorption cross sections for $l=0,1,2$ and different choices of parameter $X$}
	\label{fig:Psigma}
\end{figure}

\begin{figure}[h]
	\centering
	\includegraphics[width=80mm]{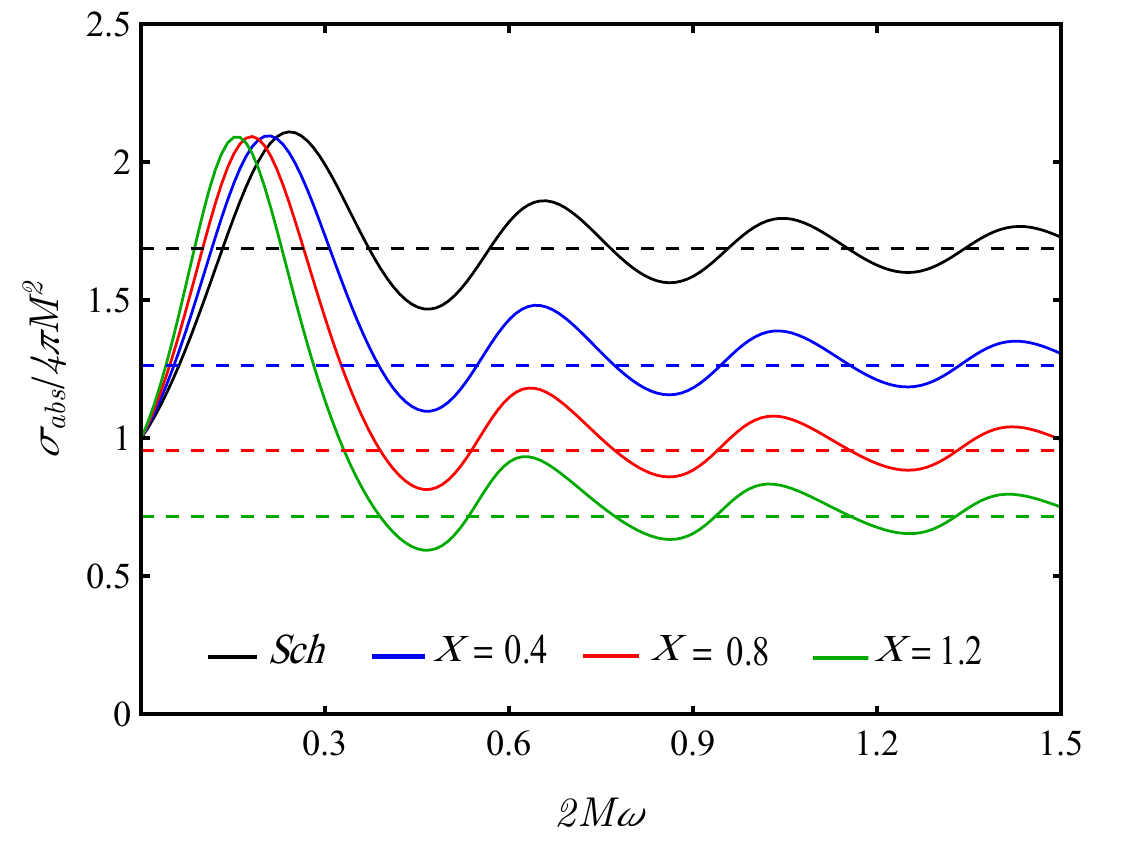} 
	\caption{Total absorption cross sections up to $l=6$ for different choices of parameter $X$. The geometric cross section is represented by dashed horizontal lines.}
	\label{fig:Tsigma}
\end{figure}

Fig. \ref{fig:Psigma} demonstrates the partial absorption cross sections for $X=0.4, 0.8$, and $1.2$ and also the Schwarzschild case $X=0$. Naturally, the partial absorption goes up as the parameter $X$ increases. These results align with the behavior of effective potential in Fig. \ref{fig:Veff}, as we expect when the height of effective potential decreases with a higher value of $X$, the absorption cross--section goes up by the observer. 
Fig. \ref{fig:Tsigma} shows the total absorption cross section obtained by adding up the $l$ modes from $l=0$ to $l=6$. We can notice that the first peak $(l=0)$ in Fig. \ref{fig:Tsigma} undergoes significant changes for different values of $X$ while for higher values of $l$, the peaks remain almost unchanged.


8\subsubsection{High frequency regime}

This section will investigate the high--frequency limit of absorption cross section, also called geometric cross section $\sigma_{geo}$. Due to the scaling of the radial coordinate with $\beta$, we obtain
\begin{equation}\label{eq:geocross}
    \sigma_{geo}=\frac{\pi b_c^2}{\beta},
\end{equation}
where $b_c$ is the critical impact parameter of the light ray incoming close to the black hole vicinity. To find this parameter, we need to explore the null geodesics, which has been done in our previous work \cite{Filho:2022yrk,araujo2024gravitational}. In summary, the light ray's equation of motion is found from Lagrangian $2\mathcal{L}=-f(r)\dot{t}^2+f(r)^{-1}\beta\dot{r}^2+r^2\dot{\phi}^2=0$. By considering two constants of motion energy $E$ and angular momentum $L$, we arrive at the equation of motion for light ray
\begin{equation}
\beta\dot{r}^2+\tilde{V}_{eff}=0,
\end{equation}
where the effective potential is given by
\begin{equation}
    \tilde{V}_{eff}=E^2-\frac{f(r)}{r^2}L^2.
\end{equation}

The light--like circular orbit, denoted by $r=r_{c}$, can be obtained by computing the expression 
 \begin{align}
\tilde{V}_{eff}&=0,\\
\frac{\mathrm{d}\tilde{V}_{eff}}{\mathrm{d}r}&=0.
\end{align}
Each equation leads to the following expressions for critical orbits, respectively.
\begin{align}\label{cr}
    &2f(r_{c})-r_{c}f'(r_{c})=0,\\ \label{impact}
    &b_c=\frac{r_{c}^2}{f(r_{c})},
\end{align}
where we have defined $b=L/E$. We note here that, since the lapse function $f(r)$ is the same as in the Schwarzschild spacetime, there are no changes in the values of $r_c$ and $b_c$, being given by $3M$ and $3\sqrt{3}M$, respectively. Nonetheless, due to the changes in the scaling factor, the geometrical cross--section is given by Eq.~\eqref{eq:geocross}, which results in $\sigma_{geo}=\sigma_{GR}/\beta$.

In Fig. \ref{fig:Tsigma}, the graph shows the total absorption cross section up to $l = 6$ for various values of $X$. According to the high frequency approximation, the absorption cross section should approach the geometrical cross section. The dotted horizontal lines in Fig. \ref{fig:Tsigma} represent the geometrical cross section. It is notable that as the frequency increases, it tends to the same value as the geometrical cross section, which is $\sigma_{geo}$.


\section{Greybody Factor}\label{Tbound}
Another crucial aspect of the scattering is the quantity related to the spectrum observed by an observer at infinity. The transmission amplitude, called the greybody factor can determine the probability of the \textit{Hawking} radiation that reaches spatial infinity. Different methods are developed to calculate the greybody factor. The WKB method has been applied in Ref. \cite{araujo2024gravitational} to investigate the effect of the bumblebee gravity parameter on the greybody factor. We utilize another method introduced in Ref.\cite{visser1999some,boonserm2008bounding,sakalli2022topical}. In this approach a general semi--analytical bound for the greybody factor is given by
\begin{equation}
{T_b} \ge {\mathop{\rm sech}\nolimits} ^2 \left(\int\limits_{ - \infty }^{ + \infty } {\mathcal{G} \rm{d}}r^{*}\right),
\end{equation}
\begin{equation}
	\mathcal{G} = \frac{{\sqrt {{{(h')}^2} + {{({\omega ^2} - {V_{eff}} - {h^2})}^2}} }}{{2h}}.
\end{equation}

Here $h$ is a positive function that satisfies the conditions $h(r^{*})>1$ and $h(-\infty)=h(+\infty)=\omega$. Substituting the $h(r)$, the effective potential from Eq. \eqref{Veff}  we arrive at
\begin{equation}
	T_b \ge {\mathop{\rm sech}\nolimits} {\left( {\frac{1}{{2\omega }}\left( {l(l + 1) + \frac{{\sqrt {{{\left( {1 - \frac{X}{4}} \right)}^3}} }}{{2\sqrt {1 + \frac{{3X}}{4}} }}} \right)} \right) ^2}.
\end{equation}

\begin{figure}[h]
	\centering
	\includegraphics[width=80mm]{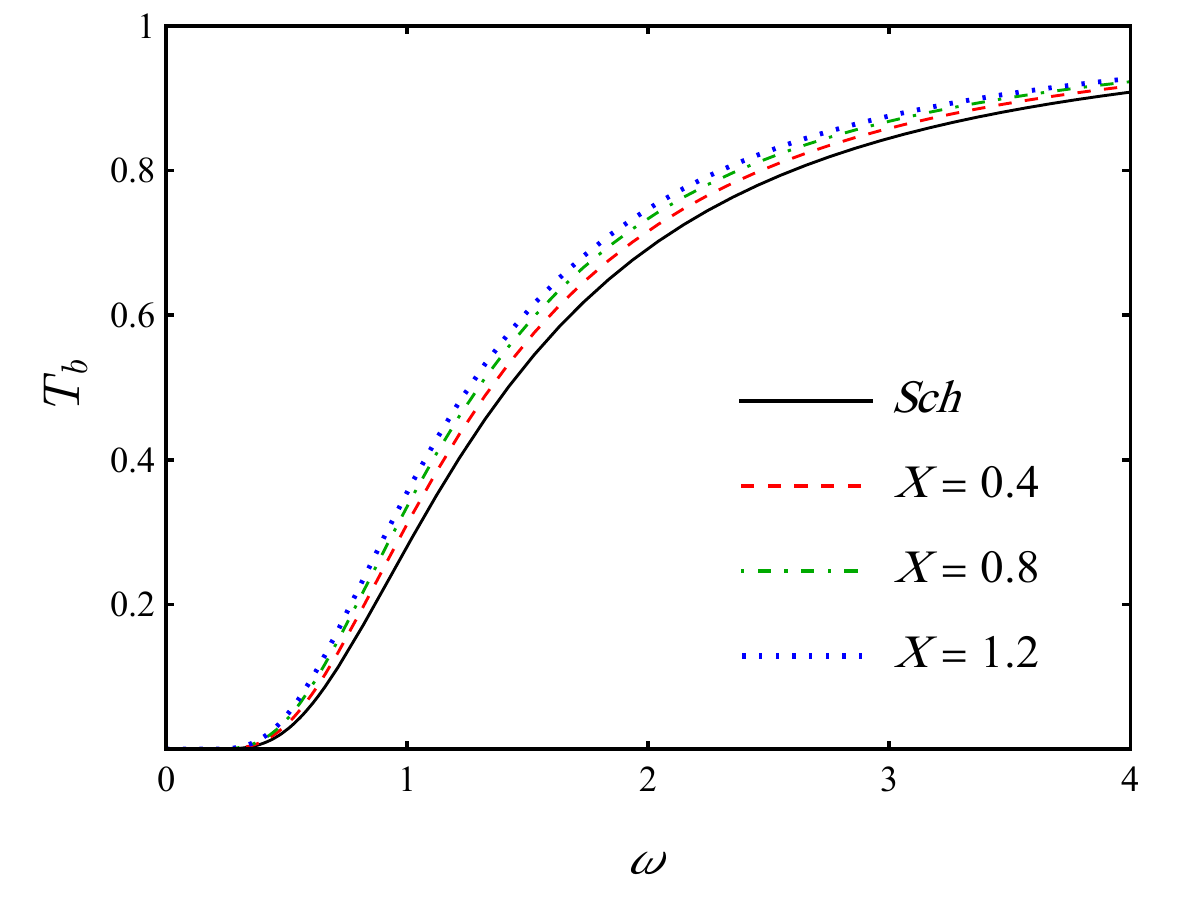}\\
	\caption{The representation of the greybody factors as a function of $\omega$ for different values of Lorentz--violating parameter $X$.}
	\label{fig:Tb}
\end{figure}

Fig. \ref{fig:Tb} illustrate that the greybody bound value is zero when the frequency is at its minimum and becomes $1$ when the frequency is high enough.

This implies that when the frequency is at a lower level, the reflected wave is total. However, as the frequency rises, a part of the wave can pass through the potential barrier due to the tunneling effect. When the frequency reaches a certain critical value, the wave is no longer reflected. Additionally, it is noticeable that the greybody bound increases with larger values of $X$. Therefore, when $X \neq 0$, due to the bumblebee gravity effects, the incident wave experiences significant scattering rather than the case of the Schwarzschild black hole. As the value of $X$ increases, the potential barrier's height decreases, which in turn, enables a higher probability of transmission to the incident wave.

\section{Quasinormal modes}\label{QNMs}

\subsection{The scaling of spherical modes and long--living modes}

One of the most accurate methods to compute the quasinormal modes relies on a recursive expansion that leads to a continued fraction (CF) expression. In BHs, this method was due to Leaver~\cite{leaver1985analytic}, and extended later on to many different cases and scenarios~\cite{Pani:2013pma}. In this section, we show how to apply the CF method in our case and display some results concerning the scalar QNMs. We have compared the results of this section against a direct integration method obtaining good agreement for the fundamental mode.

We start by writing an series expression for the scalar field that satisfies both boundary conditions required in the QNM computation. Therefore, we have
\begin{equation}
    \psi(r)= \left(\frac{r}{2M}-1\right)^{\rho } \left(\frac{r}{2M}\right)^{-2 \rho } \exp [-\rho  (r/2M-1)]\sum_{n=0}^N a_n \left(\frac{r-2M}{r}\right)^n,
\end{equation}
where $\rho=-i\sqrt{\beta}\omega$. By inserting the above expansion in the scalar field differential equation, we find the following three--term recurrence relation
\begin{align}
   & \alpha_n a_{n+1}+\beta_na_n+\gamma_n a_{n-1}=0, ~{\rm with}~ n\geq1,\\
   & \alpha_0 a_{1}+\beta_0a_0=0,
\end{align}
where
\begin{align}
    \alpha_n&=(n+1) \left(-4 i \sqrt{\beta } M \omega +n+1\right),\\
    \sigma_n&=32 \beta  M^2 \omega ^2+8 i \sqrt{\beta } M \omega +16 i \sqrt{\beta } M n \omega -2 n (n+1)-\beta \l  (l +1)-1,\\
    \gamma_n&=\left(n-4 i \sqrt{\beta } M \omega \right)^2.
\end{align}
We can solve the above recurrence relation, leading to the following algebraic equation
\begin{equation}\label{eq:CF}
    \frac{\sigma_0}{\alpha_0}-\frac{\gamma_1}{\sigma_1-\frac{\alpha_1\gamma_2}{\sigma_2-\frac{\alpha_2\gamma_3}{\sigma_3-...}}}=0,
\end{equation}
which is satisfied for given values of $\omega$ corresponding to the QNM frequencies. Note that the above expression must be truncated at a high enough value of $N$ to ensure the numerical accuracy of the modes. In our case, we consider $N={\cal O}(10^3)$, but the exact number depends on the value of $X$.

Using the abovementioned procedure, we computed the fundamental scalar perturbations and the first two overtones QNMs. One way to parameterize the departure from GR is with the normalized deviation, i.e.,
\begin{equation}
\delta=\frac{\omega_{R,I}}{\omega_{R,I|GR}}-1,
\end{equation}
where $\omega_{R,I|GR}$ is the standard mode for the Schwarzschild case. We shall use this definition to investigate some of the behavior of the modes.

Before entering in details, it is worth to note that the $l=0$ case can be computed \textit{analytically}, in terms of the GR mode. Given the expression for the scalar field equation, plunging $l=0$ we find that the GR case can be recovered by making the rescaling $\omega\to \omega/\beta^{1/2}$. Therefore, the $l=0$ QNM are given by
\begin{equation}
    \omega=\beta^{-1/2}\omega_{GR}~~~(l=0).
\end{equation}
Therefore, as $\beta\to\infty$ ($X\to4$), we have that both real and imaginary part of the $l=0$ modes -- fundamental and overtones -- goes to zero in this limit. We have also checked the above against numerical computation, obtaining excellent agreement.

\begin{figure}
    \centering
\includegraphics[width=0.5\linewidth]{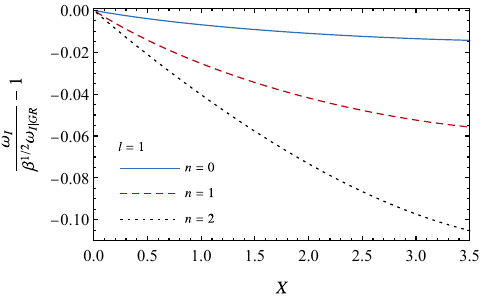}\includegraphics[width=0.5\linewidth]{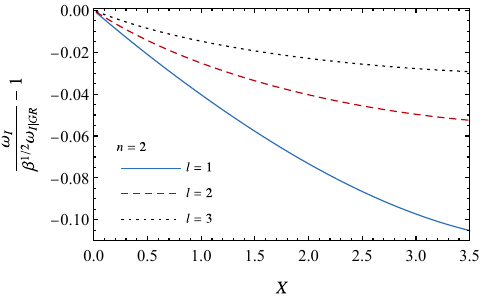}
    \caption{Normalized deviation from expression \eqref{eq:dev} of the imaginary part of the modes as a function of $X$, considering $l=1,2,3$ up to the second overtone. Deviations from Eq.~ \eqref{eq:dev} are more evident for higher overtones (left panel) and lower $l$ (right panel).}
    \label{fig:devim}
\end{figure}

\begin{figure}
    \centering
    \includegraphics[width=1\linewidth]{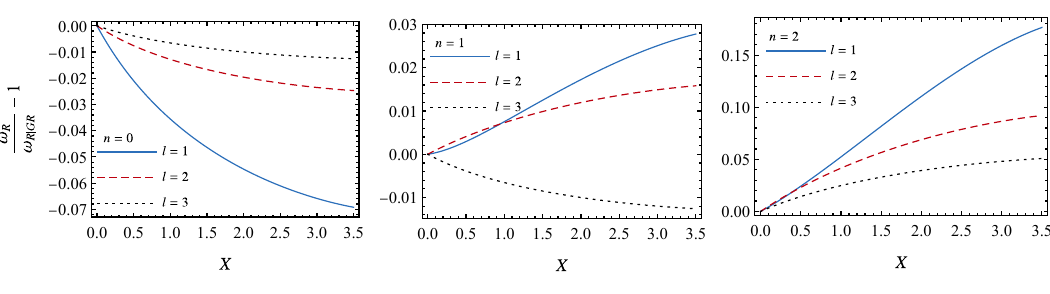}
    \caption{Normalized deviation from GR of the real part of the modes as function of $X$.}
    \label{fig:real}
\end{figure}

For higher $l$'s, there is no clear correspondence between the GR and bumblebee modes; therefore, they must be computed numerically. By computing the modes through the CF expression~\eqref{eq:CF}, we found a non--trivial dependence of the real part as a function of $X$. Despite this, we still find (numerically) that the imaginary part for all $l$ and the first overtones behave approximately as
\begin{equation}\label{eq:dev}
\omega_I\approx\beta^{-1/2}\omega_{I|GR}.
\end{equation}
In Fig.~\ref{fig:devim} we show the normalized deviation of the numerically computed imaginary part against \eqref{eq:dev} for the $l=0,1,2$, considering $n=0,1,2$. We see that deviations from \eqref{eq:dev} are more evident for lower $l$ and higher $n$. In the cases displayed in Fig.~\ref{fig:devim}, we see deviations of up to $\sim 10\%$. This indicates the existence of long--living modes in the limit $\beta\to\infty$ ($X\to4$).
For the real part, as mentioned above, there is no clear evident pattern for the behavior of the modes as a function of $X$. In Fig.~\ref{fig:real} we show the deviation of the real part of the modes as function of $X$ for $n=0,1,2$ para $l=1,2,3$. We see that while for $n=0$ the mode decreases as compared to GR, the $n=1$ behavior depends on $l$, and for $n=2$ the frequency increases. Note that the bigger deviations occur, once again, for higher overtones and lower $l$.

\subsection{Relation between Shadow radius and quasinormal mode}

As discussed in previous sections, quasinormal modes are characterized by a complex number $\omega=\omega_R-i\omega_I$. Initially, it was shown that the real and imaginary parts of the QNMs are related to the angular
velocity and Lyapunov exponent of unstable circular null
geodesics by Cardoso et al.\cite{cardoso2009geodesic}
\begin{equation}\label{eikon0}
	\omega  = l\Omega  - i(n + \frac{1}{2})|\lambda |.
\end{equation}
where $\Omega$ and $\lambda$ denote the the angular
velocity and Lyapunov exponent of the photon sphere, respectively. Also, $n$ is the overtone number and $l$ is the angular quantum number called multipole number.
Then, a connection between black hole quasinormal modes in the eikonal limit and lensing in the strong deflection limit has been discussed by Stefanoc et al. \cite{stefanov2010connection,wei2011relationship,wei2014establishing}. Following these results, Jusufi \cite{jusufi2020connection,liu2020shadow,jusufi2020quasinormal} has pointed out that the real part of QNMs has an inverse relation with the shadow radius as 
\begin{equation}\label{eikon1}
	{\omega _R} = \mathop {\lim\limits_{l \gg 1}  \frac{l}{{{R_{\text{sh}}}}}}.
\end{equation}
The inverse relation between shadow radius and the real part of quasinormal mode is valid for the static spherical spacetime and accurate for the eikonal limit $l \gg 1$ and, in ref. \cite{cuadros2020analytical}, Eq. \eqref{eikon1} has been improved to the sub--leading regime to half of its value as
\begin{equation}\label{eikon2}
	{\omega _R} = \mathop {\lim\limits_{l \gg 1}  \frac{l+\frac{1}{2}}{{{R_{\text{sh}}}}}}.
\end{equation}

\begin{figure}[h]
	\centering
	\includegraphics[width=80mm]{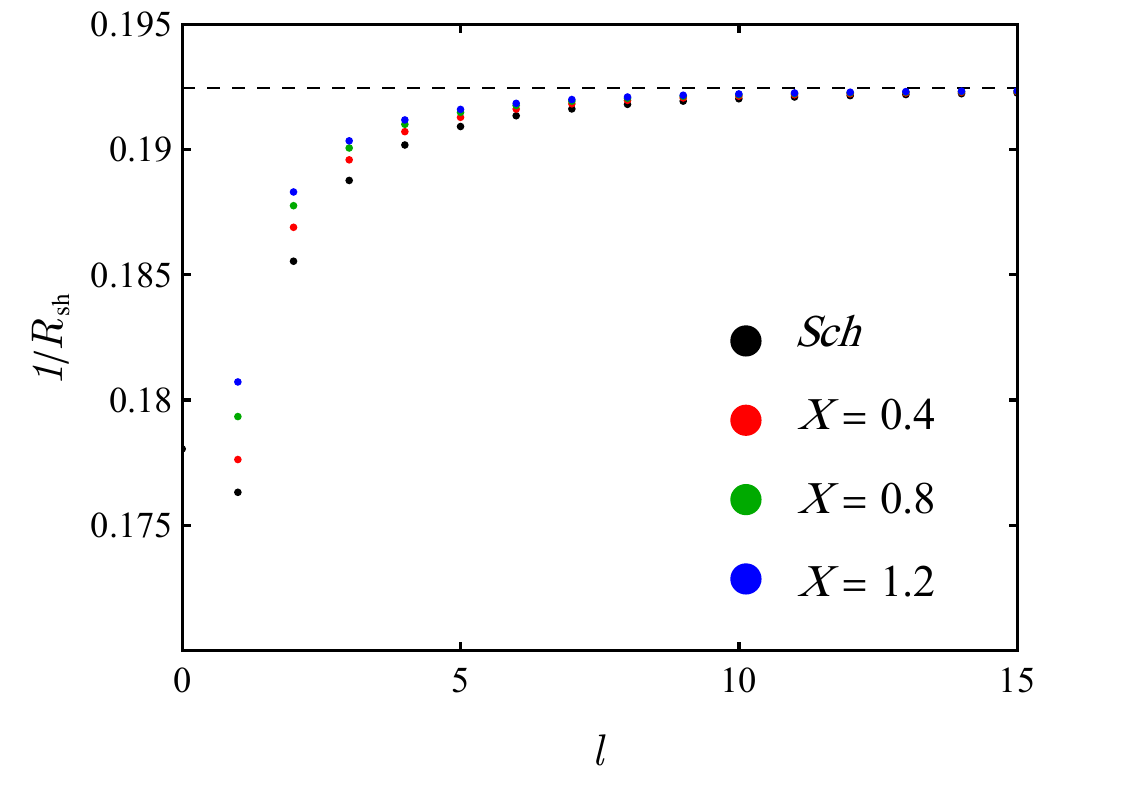} 
	\caption{The Dots show the values of $\omega_{R}\left(l+\frac{1}{2}\right)^{-1}$ and dashed lines indicate the inverse of shadow radius ($R_{\text{sh}}^{-1}$). Sixth-order WKB method applied to calculate $\omega_R$ for $M=1$ and $n=1$. }
	\label{fig:eikonal}
\end{figure}

This relation has been investigated for various black holes \cite{ma2023shadow,liu2020shadow}.
Following \cite{perlick2022calculating},the critical radius can be obtained by solving the Eq. \eqref{cr}, and the critical radius will be $r_{c}=3M$. Now, by considering the  observer radius to be infinity, the shadow radius can be found by 
\begin{equation}\label{Rsh1}
	R_{\text{sh}} =\sqrt {\frac{{r_{c}}^2}{{f({r_{c}})}}},
\end{equation}
which leads to $R_{\text{sh}}=3\sqrt{3}M$ for the related spacetime in Eq. \eqref{dsbar}.
The real part of the quasinormal frequencies for eikonal limits are obtained based on 6th--order WKB method. In Fig. \ref{fig:eikonal}, the real part of QNMs divided by $l+\frac{1}{2}$ plotted against the multipole values in the range of $1 < l < 120$. The inverse of the shadow radius is shown by dashed lines. This plot illustrates that with increasing multipole values, there is a convergence of quasinormal frequency values towards the inverse of the shadow radius which is expected from Eq. \eqref{eikon0}

\section{Emission rate}\label{EMR}

The inside of black holes experiences quantum fluctuations that result in the creation and annihilation of numerous particles near the horizon. Particles with positive energy can escape from the black hole through a process called tunneling. This causes the black hole to gradually evaporate over time, a phenomenon known as \textit{Hawking} radiation. From the perspective of a distant observer, the black hole shadow represents a high--energy absorption cross--section, which has been shown to be around a constant value $\sigma_{lim}$, and following Ref. \cite{decanini2011universality,papnoi2022rotating} the energy emission rates can be obtained as
\begin{equation}\label{emission}
	\frac{{{\mathrm{d}^2}E}}{{\mathrm{d}\omega \mathrm{d}t}} = \frac{{2{\pi ^2}\sigma_{lim}}}{{{e^{\frac{\omega }{T_H}}} - 1}} {\omega ^3},
\end{equation}
where $\omega$ denotes the frequency of a photon, and $T_H$ symbolizes the \textit{Hawking} temperature. The limiting constant value $\sigma_{lim}$ connected to the shadow radius as $\sigma_{lim}\approx \pi R_{\text{sh}}^2$. Substituting the shadow radius in Eq. \ref{Rsh1} and \textit{Hawking} temperature, the rate of energy emission is investigated for various values of the bumblebee parameter.

\begin{figure}[h]
	\centering
	\includegraphics[width=80mm]{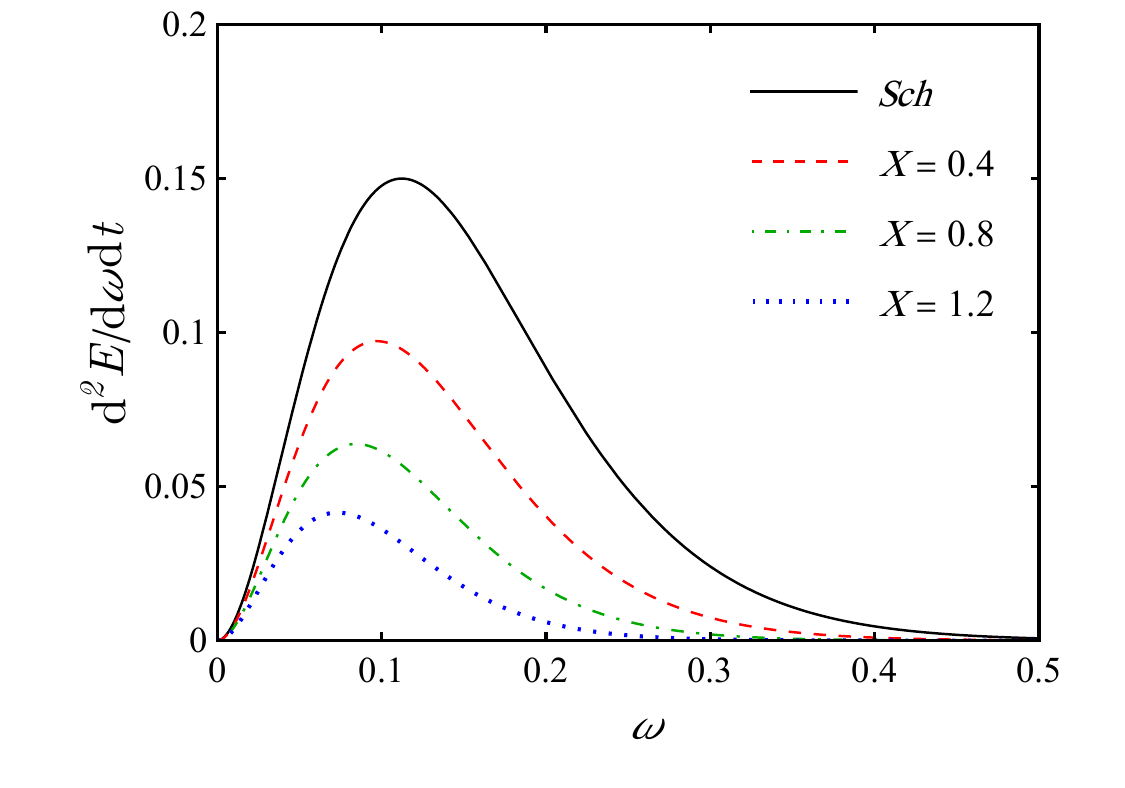} 
	\caption{The behavior of energy emission rate against the frequency, for $M = 1$.}
	\label{fig:Emission}
\end{figure}
In Fig. \ref{fig:Emission}, the energy emission rate against the frequency is represented for $M = 1$ and different values of the $X$.
It can be observed that the bigger values of $X$ conduct a lower emission rate and shift the maximum of plots to a smaller frequency. This indicates that the higher bumblebee parameter corresponds to the slower evaporation process in the black hole.


\section{Conclusion}
\label{conclusion}
In this work, we sought a better understanding of the unique characteristics and consequences of Schwarzschild-like black holes within the framework of metric-affine bumblebee gravity.
We showed that, similar to the Schwarzschild black hole, the Kretschmann scalar is singular only in $r = 0$. We investigate wave-like phenomena using a test scalar field.

We computed the absorption cross-section using numerical partial wave computations, showing that the value goes to the horizon area for low frequencies. For higher frequencies, we showed that the capture cross-section for null geodesics with a scaling factor agrees with the partial wave analysis.

The greybody bound has been investigated and it was discovered that the greybody bound increases with larger $X$. Consequently, propagating waves can greatly pass through the potential barrier for bumblebee spacetime with large $X$.

By using a continued fraction expansion for the scalar field, we were able to compute the quasinormal mode frequencies in a wide range of the parameter space, showing some characteristics of bumblebee black holes, such as the possibility of having low imaginary modes, which ring longer in dynamical scenarios.

Then by exploring the eikonal limit, we noted that the correspondence between the real part of QNMs and the shadow radius is also satisfied in this modified gravity theorem. Finally, through analysis of the energy emission rate, we have discovered that the bumblebee metric affine parameter has a diminishing effect on the emission rate, demonstrating that a larger bumblebee parameter will result in a slower evaporation process for the black hole in this framework.

\bibliography{main}

\begin{thebibliography}{100}

\bibitem{STR1}
Simon Judes and Matt Visser.
\newblock Conservation laws in ``doubly special relativity''.
\newblock {\em Phys. Rev. D}, 68:045001, 2003.

\bibitem{STR2}
H.~P. Robertson.
\newblock Postulate versus observation in the special theory of relativity.
\newblock {\em Rev. Mod. Phys.}, 21:378--382, 1949.

\bibitem{STR3}
Robert~C. Myers and Maxim Pospelov.
\newblock Ultraviolet modifications of dispersion relations in effective field theory.
\newblock {\em Phys. Rev. Lett.}, 90:211601, 2003.

\bibitem{STR4}
O.~Bertolami and J.~G. Rosa.
\newblock Bounds on cubic lorentz-violating terms in the fermionic dispersion relation.
\newblock {\em Phys. Rev. D}, 71:097901, 2005.

\bibitem{STR5}
C.~M. Reyes, L.~F. Urrutia, and J.~D. Vergara.
\newblock Quantization of the myers-pospelov model: The photon sector interacting with standard fermions as a perturbation of qed.
\newblock {\em Phys. Rev. D}, 78:125011, 2008.

\bibitem{STR6}
David Mattingly.
\newblock Have we tested lorentz invariance enough?
\newblock {\em arXiv preprint arXiv:0802.1561}, 2008.

\bibitem{STR7}
G.I. Rubtsov, P.S. Satunin, and S.M. Sibiryakov.
\newblock The influence of lorentz violation on uhe photon detection.
\newblock {\em CPT and Lorentz Symmetry}, page 192–195, 2014.

\bibitem{liberati2013}
Stefano Liberati.
\newblock Tests of lorentz invariance: a 2013 update.
\newblock {\em Classical and Quantum Gravity}, 30(13):133001, 2013.

\bibitem{tasson2014}
Jay~D Tasson.
\newblock What do we know about lorentz invariance?
\newblock {\em Reports on Progress in Physics}, 77(6):062901, 2014.

\bibitem{hees2016}
Aur{\'e}lien Hees, Quentin~G Bailey, Adrien Bourgoin, Pihan-Le Bars, Christine Guerlin, Le~Poncin-Lafitte, et~al.
\newblock Tests of lorentz symmetry in the gravitational sector.
\newblock {\em Universe}, 2(4):30, 2016.

\bibitem{rovelli2004}
Carlo Rovelli.
\newblock {\em Quantum gravity}.
\newblock Cambridge university press, 2004.

\bibitem{New1}
V.~Alan Kosteleck\'y and Stuart Samuel.
\newblock Spontaneous breaking of lorentz symmetry in string theory.
\newblock {\em Phys. Rev. D}, 39:683--685, 1989.

\bibitem{New2}
V.~Alan Kosteleck\'y and Stuart Samuel.
\newblock Phenomenological gravitational constraints on strings and higher-dimensional theories.
\newblock {\em Phys. Rev. Lett.}, 63:224--227, 1989.

\bibitem{New3}
V.~Alan Kosteleck\'y and Stuart Samuel.
\newblock Gravitational phenomenology in higher-dimensional theories and strings.
\newblock {\em Phys. Rev. D}, 40:1886--1903, 1989.

\bibitem{New4}
V.~Alan Kostelecký and Robertus Potting.
\newblock Cpt and strings.
\newblock {\em Nuclear Physics B}, 359(2):545 -- 570, 1991.

\bibitem{New5}
V.~Alan Kosteleck\'y and Robertus Potting.
\newblock Cpt, strings, and meson factories.
\newblock {\em Phys. Rev. D}, 51:3923--3935, 1995.

\bibitem{New6}
Rodolfo Gambini and Jorge Pullin.
\newblock Nonstandard optics from quantum space-time.
\newblock {\em Phys. Rev. D}, 59:124021, 1999.

\bibitem{New7}
Martin Bojowald, Hugo~A. Morales-T\'ecotl, and Hanno Sahlmann.
\newblock Loop quantum gravity phenomenology and the issue of lorentz invariance.
\newblock {\em Phys. Rev. D}, 71:084012, 2005.

\bibitem{New8}
Giovanni Amelino-Camelia and Shahn Majid.
\newblock Waves on noncommutative space--time and gamma-ray bursts.
\newblock {\em International Journal of Modern Physics A}, 15(27):4301--4323, 2000.

\bibitem{New9}
Sean~M. Carroll, Jeffrey~A. Harvey, V.~Alan Kosteleck\'y, Charles~D. Lane, and Takemi Okamoto.
\newblock Noncommutative field theory and lorentz violation.
\newblock {\em Phys. Rev. Lett.}, 87:141601, 2001.

\bibitem{Modesto:2011kw}
Leonardo Modesto.
\newblock {Super-renormalizable Quantum Gravity}.
\newblock {\em Phys. Rev. D}, 86:044005, 2012.

\bibitem{Nascimento:2021bzb}
J.~R. Nascimento, A.~Yu. Petrov, and P.~J. Porf\'\i{}rio.
\newblock {Causal G\"odel-type metrics in non-local gravity theories}.
\newblock {\em Eur. Phys. J. C}, 81(9):815, 2021.

\bibitem{New10}
Frans~R. Klinkhamer and Christian Rupp.
\newblock Spacetime foam, cpt anomaly, and photon propagation.
\newblock {\em Phys. Rev. D}, 70:045020, 2004.

\bibitem{New11}
S.~Bernadotte and F.~R. Klinkhamer.
\newblock Bounds on length scales of classical spacetime foam models.
\newblock {\em Phys. Rev. D}, 75:024028, 2007.

\bibitem{New12}
F.R. Klinkhamer.
\newblock Z-string global gauge anomaly and lorentz non-invariance.
\newblock {\em Nuclear Physics B}, 535(1):233 -- 241, 1998.

\bibitem{New13}
F.R. Klinkhamer.
\newblock A cpt anomaly.
\newblock {\em Nuclear Physics B}, 578(1):277 -- 289, 2000.

\bibitem{New14}
F.R. Klinkhamer and J.~Schimmel.
\newblock Cpt anomaly: a rigorous result in four dimensions.
\newblock {\em Nuclear Physics B}, 639(1):241 -- 262, 2002.

\bibitem{New15}
K.J.B. Ghosh and F.R. Klinkhamer.
\newblock Anomalous lorentz and cpt violation from a local chern–simons-like term in the effective gauge-field action.
\newblock {\em Nuclear Physics B}, 926:335 -- 369, 2018.

\bibitem{New16}
Petr Ho\ifmmode~\check{r}\else \v{r}\fi{}ava.
\newblock Quantum gravity at a lifshitz point.
\newblock {\em Phys. Rev. D}, 79:084008, 2009.

\bibitem{sv1}
Guido Cognola, Ratbay Myrzakulov, Lorenzo Sebastiani, Sunny Vagnozzi, and Sergio Zerbini.
\newblock Covariant ho{\v{r}}ava-like and mimetic horndeski gravity: cosmological solutions and perturbations.
\newblock {\em Classical and quantum gravity}, 33(22):225014, 2016.

\bibitem{sv2}
Alessandro Casalino, Massimiliano Rinaldi, Lorenzo Sebastiani, and Sunny Vagnozzi.
\newblock Alive and well: mimetic gravity and a higher-order extension in light of gw170817.
\newblock {\em Classical and Quantum Gravity}, 36(1):017001, 2018.

\bibitem{kostelecky2011data}
V~Alan Kosteleck{\`y} and Neil Russell.
\newblock Data tables for lorentz and c p t violation.
\newblock {\em Reviews of Modern Physics}, 83(1):11, 2011.

\bibitem{colladay2004statistical}
Don Colladay and Patrick McDonald.
\newblock Statistical mechanics and lorentz violation.
\newblock {\em Physical Review D}, 70(12):125007, 2004.

\bibitem{aa2021lorentz}
A.~A Ara{\'u}jo~Filho.
\newblock Lorentz-violating scenarios in a thermal reservoir.
\newblock {\em The European Physical Journal Plus}, 136(4):1--14, 2021.

\bibitem{araujo2021thermodynamic}
A.~A Ara{\'u}jo~Filho and R.~V Maluf.
\newblock Thermodynamic properties in higher-derivative electrodynamics.
\newblock {\em Brazilian Journal of Physics}, 51(3):820--830, 2021.

\bibitem{anacleto2018lorentz}
MA~Anacleto, FA~Brito, E~Maciel, A~Mohammadi, E~Passos, WO~Santos, and JRL Santos.
\newblock Lorentz-violating dimension-five operator contribution to the black body radiation.
\newblock {\em Physics Letters B}, 785:191--196, 2018.

\bibitem{casana2008lorentz}
Rodolfo Casana, Manoel~M Ferreira~Jr, and Josberg~S Rodrigues.
\newblock Lorentz-violating contributions of the carroll-field-jackiw model to the cmb anisotropy.
\newblock {\em Physical Review D}, 78(12):125013, 2008.

\bibitem{casana2009finite}
Rodolfo Casana, Manoel~M Ferreira~Jr, Josberg~S Rodrigues, and Madson~RO Silva.
\newblock Finite temperature behavior of the c p t-even and parity-even electrodynamics of the standard model extension.
\newblock {\em Physical Review D}, 80(8):085026, 2009.

\bibitem{araujo2021higher}
A.~A Ara{\'u}jo~Filho and A~Yu Petrov.
\newblock Higher-derivative lorentz-breaking dispersion relations: a thermal description.
\newblock {\em The European Physical Journal C}, 81(9):843, 2021.

\bibitem{aguirre2021lorentz}
AR~Aguirre, G~Flores-Hidalgo, RG~Rana, and ES~Souza.
\newblock The lorentz-violating real scalar field at thermal equilibrium.
\newblock {\em The European Physical Journal C}, 81(5):459, 2021.

\bibitem{Mariz:2011ed}
T.~Mariz, J.~R. Nascimento, and A.~Yu. Petrov.
\newblock {On the perturbative generation of the higher-derivative Lorentz-breaking terms}.
\newblock {\em Phys. Rev. D}, 85:125003, 2012.

\bibitem{reis2021thermal}
J.~A. A.~S. Reis et~al.
\newblock Thermal aspects of interacting quantum gases in lorentz-violating scenarios.
\newblock {\em The European Physical Journal Plus}, 136(3):310, 2021.

\bibitem{petrov2021bouncing2}
A.~A Ara{\'u}jo~Filho and A~Yu Petrov.
\newblock Bouncing universe in a heat bath.
\newblock {\em International Journal of Modern Physics A}, 36(34 \& 35):2150242, 2021.

\bibitem{furtado2023thermal}
J~Furtado, H~Hassanabadi, JAAS Reis, et~al.
\newblock Thermal analysis of photon-like particles in rainbow gravity.
\newblock {\em arXiv preprint arXiv:2305.08587}, 2023.

\bibitem{aaa2021thermodynamics}
A.~A Ara{\'u}jo~Filho.
\newblock Thermodynamics of massless particles in curved spacetime.
\newblock {\em arXiv preprint arXiv:2201.00066}, 2021.

\bibitem{CFJ}
Sean~M Carroll, George~B Field, and Roman Jackiw.
\newblock Limits on a lorentz-and parity-violating modification of electrodynamics.
\newblock {\em Physical Review D}, 41(4):1231, 1990.

\bibitem{aether}
Sean~M Carroll and Heywood Tam.
\newblock Aether compactification.
\newblock {\em Physical Review D}, 78(4):044047, 2008.

\bibitem{Gomes:2009ch}
M.~Gomes, J.~R. Nascimento, A.~Yu. Petrov, and A.~J. da~Silva.
\newblock {On the aether-like Lorentz-breaking actions}.
\newblock {\em Phys. Rev. D}, 81:045018, 2010.

\bibitem{colladay1998lorentz}
Don Colladay and V~Alan Kosteleck{\`y}.
\newblock Lorentz-violating extension of the standard model.
\newblock {\em Physical Review D}, 58(11):116002, 1998.

\bibitem{5}
V~Alan Kosteleck{\`y}.
\newblock Gravity, lorentz violation, and the standard model.
\newblock {\em Physical Review D}, 69(10):105009, 2004.

\bibitem{kostelecky2021backgrounds}
V~Alan Kosteleck{\`y} and Zonghao Li.
\newblock Backgrounds in gravitational effective field theory.
\newblock {\em Physical Review D}, 103(2):024059, 2021.

\bibitem{bluhm2003probing}
Robert Bluhm, V~Alan Kosteleck{\`y}, Charles~D Lane, and Neil Russell.
\newblock Probing lorentz and cpt violation with space-based experiments.
\newblock {\em Physical Review D}, 68(12):125008, 2003.

\bibitem{bluhm2005spontaneous}
Robert Bluhm and V~Alan Kosteleck{\`y}.
\newblock Spontaneous lorentz violation, nambu-goldstone modes, and gravity.
\newblock {\em Physical Review D—Particles, Fields, Gravitation, and Cosmology}, 71(6):065008, 2005.

\bibitem{bluhm2008spontaneous}
Robert Bluhm, Shu-Hong Fung, and V~Alan Kosteleck{\`y}.
\newblock Spontaneous lorentz and diffeomorphism violation, massive modes, and gravity.
\newblock {\em Physical Review D—Particles, Fields, Gravitation, and Cosmology}, 77(6):065020, 2008.

\bibitem{bluhm2021gravity}
Robert Bluhm and Yumu Yang.
\newblock Gravity with explicit diffeomorphism breaking.
\newblock {\em Symmetry}, 13(4):660, 2021.

\bibitem{bluhm2023spontaneous}
Robert Bluhm and Yu~Zhi.
\newblock Spontaneous and explicit spacetime symmetry breaking in einstein--cartan theory with background fields.
\newblock {\em Symmetry}, 16(1):25, 2023.

\bibitem{kostelecky2011matter}
V~Alan Kosteleck{\`y} and Jay~D Tasson.
\newblock Matter-gravity couplings and lorentz violation.
\newblock {\em Physical Review D—Particles, Fields, Gravitation, and Cosmology}, 83(1):016013, 2011.

\bibitem{6}
O~Bertolami and J~Paramos.
\newblock Vacuum solutions of a gravity model with vector-induced spontaneous lorentz symmetry breaking.
\newblock {\em Physical Review D}, 72(4):044001, 2005.

\bibitem{7}
R~Casana, A~Cavalcante, FP~Poulis, and EB~Santos.
\newblock Exact schwarzschild-like solution in a bumblebee gravity model.
\newblock {\em Physical Review D}, 97(10):104001, 2018.

\bibitem{8}
AF~Santos, WDR Jesus, JR~Nascimento, and A~Yu Petrov.
\newblock G{\"o}del solution in the bumblebee gravity.
\newblock {\em Modern Physics Letters A}, 30(02):1550011, 2015.

\bibitem{9}
WDR Jesus and AF~Santos.
\newblock G{\"o}del-type universes in bumblebee gravity.
\newblock {\em International Journal of Modern Physics A}, 35(09):2050050, 2020.

\bibitem{10}
WDR Jesus and AF~Santos.
\newblock Ricci dark energy in bumblebee gravity model.
\newblock {\em Modern Physics Letters A}, 34(22):1950171, 2019.

\bibitem{11}
RV~Maluf and Juliano~CS Neves.
\newblock Black holes with a cosmological constant in bumblebee gravity.
\newblock {\em Physical Review D}, 103(4):044002, 2021.

\bibitem{12}
RV~Maluf and Juliano~CS Neves.
\newblock Black holes with a cosmological constant in bumblebee gravity.
\newblock {\em Physical Review D}, 103(4):044002, 2021.

\bibitem{13}
Sohan~Kumar Jha, Himangshu Barman, and Anisur Rahaman.
\newblock Bumblebee gravity and particle motion in snyder noncommutative spacetime structures.
\newblock {\em Journal of Cosmology and Astroparticle Physics}, 2021(04):036, 2021.

\bibitem{14}
Rui Xu, Dicong Liang, and Lijing Shao.
\newblock Static spherical vacuum solutions in the bumblebee gravity model.
\newblock {\em Physical Review D}, 107(2):024011, 2023.

\bibitem{Maluf:2021lwh}
R.~V. Maluf and Juliano C.~S. Neves.
\newblock {Bumblebee field as a source of cosmological anisotropies}.
\newblock {\em JCAP}, 10:038, 2021.

\bibitem{KumarJha:2020ivj}
Sohan Kumar~Jha, Himangshu Barman, and Anisur Rahaman.
\newblock {Bumblebee gravity and particle motion in Snyder noncommutative spacetime structures}.
\newblock {\em JCAP}, 04:036, 2021.

\bibitem{15}
Ted Jacobson and David Mattingly.
\newblock Gravity with a dynamical preferred frame.
\newblock {\em Physical Review D}, 64(2):024028, 2001.

\bibitem{16}
Roman Jackiw and S-Y Pi.
\newblock Chern-simons modification of general relativity.
\newblock {\em Physical Review D}, 68(10):104012, 2003.

\bibitem{17}
Leila Mirzagholi, Eiichiro Komatsu, Kaloian~D Lozanov, and Yuki Watanabe.
\newblock Effects of gravitational chern-simons during axion-su (2) inflation.
\newblock {\em Journal of Cosmology and Astroparticle Physics}, 2020(06):024, 2020.

\bibitem{18}
Nicola Bartolo and Giorgio Orlando.
\newblock Parity breaking signatures from a chern-simons coupling during inflation: the case of non-gaussian gravitational waves.
\newblock {\em Journal of Cosmology and Astroparticle Physics}, 2017(07):034, 2017.

\bibitem{19}
Aindri{\'u} Conroy and Tomi Koivisto.
\newblock Parity-violating gravity and gw170817 in non-riemannian cosmology.
\newblock {\em Journal of Cosmology and Astroparticle Physics}, 2019(12):016, 2019.

\bibitem{20}
Mingzhe Li, Haomin Rao, and Dehao Zhao.
\newblock A simple parity violating gravity model without ghost instability.
\newblock {\em Journal of Cosmology and Astroparticle Physics}, 2020(11):023, 2020.

\bibitem{Rao:2023doc}
Haomin Rao and Dehao Zhao.
\newblock {Parity violating scalar-tensor model in teleparallel gravity and its cosmological application}.
\newblock {\em JHEP}, 08:070, 2023.

\bibitem{21}
PJ~Porfirio, JB~Fonseca-Neto, JR~Nascimento, A~Yu Petrov, J~Ricardo, and AF~Santos.
\newblock Chern-simons modified gravity and closed timelike curves.
\newblock {\em Physical Review D}, 94(4):044044, 2016.

\bibitem{22}
PJ~Porfirio, JB~Fonseca-Neto, JR~Nascimento, and A~Yu Petrov.
\newblock Causality aspects of the dynamical chern-simons modified gravity.
\newblock {\em Physical Review D}, 94(10):104057, 2016.

\bibitem{23}
B~Altschul, JR~Nascimento, A~Yu Petrov, and PJ~Porf{\'\i}rio.
\newblock First-order perturbations of g{\"o}del-type metrics in non-dynamical chern--simons modified gravity.
\newblock {\em Classical and Quantum Gravity}, 39(2):025002, 2021.

\bibitem{24}
Quentin~G Bailey and V~Alan Kosteleck{\`y}.
\newblock Signals for lorentz violation in post-newtonian gravity.
\newblock {\em Physical Review D}, 74(4):045001, 2006.

\bibitem{25}
Rhondale Tso and Quentin~G Bailey.
\newblock Light-bending tests of lorentz invariance.
\newblock {\em Physical Review D}, 84(8):085025, 2011.

\bibitem{26}
Aur{\'e}lien Hees, Quentin~G Bailey, Christophe Le~Poncin-Lafitte, Adrien Bourgoin, Attilio Rivoldini, Brahim Lamine, Fr{\'e}d{\'e}ric Meynadier, Christine Guerlin, and Peter Wolf.
\newblock Testing lorentz symmetry with planetary orbital dynamics.
\newblock {\em Physical Review D}, 92(6):064049, 2015.

\bibitem{LIGOScientific:2016aoc}
B.~P. Abbott et~al.
\newblock {Observation of Gravitational Waves from a Binary Black Hole Merger}.
\newblock {\em Phys. Rev. Lett.}, 116(6):061102, 2016.

\bibitem{EventHorizonTelescope:2019dse}
Kazunori Akiyama et~al.
\newblock {First M87 Event Horizon Telescope Results. I. The Shadow of the Supermassive Black Hole}.
\newblock {\em Astrophys. J. Lett.}, 875:L1, 2019.

\bibitem{EventHorizonTelescope:2022wkp}
Kazunori Akiyama et~al.
\newblock {First Sagittarius A* Event Horizon Telescope Results. I. The Shadow of the Supermassive Black Hole in the Center of the Milky Way}.
\newblock {\em Astrophys. J. Lett.}, 930(2):L12, 2022.

\bibitem{Nascimento:2021vou}
J.~R. Nascimento, A.~Yu. Petrov, and P.~J. Porf\'\i{}rio.
\newblock {Induced gravitational topological term and the Einstein-Cartan modified theory}.
\newblock {\em Phys. Rev. D}, 105(4):044053, 2022.

\bibitem{Bao}
Bao D., Chern S.-S., and Shen Z.
\newblock {\em {An Introduction to Riemann-Finsler Geometry}}.
\newblock Springer, New York, USA, 2000.

\bibitem{Foster}
Joshua Foster and Ralf Lehnert.
\newblock {Classical-physics applications for Finsler $b$ space}.
\newblock {\em Phys. Lett. B}, 746:164--170, 2015.

\bibitem{KosE}
Benjamin~R. Edwards and V.~Alan Kostelecky.
\newblock {Riemann\textendash{}Finsler geometry and Lorentz-violating scalar fields}.
\newblock {\em Phys. Lett. B}, 786:319--326, 2018.

\bibitem{Sch1}
M.~Schreck.
\newblock {Classical kinematics and Finsler structures for nonminimal Lorentz-violating fermions}.
\newblock {\em Eur. Phys. J. C}, 75(5):187, 2015.

\bibitem{CollM}
Don Colladay and Patrick McDonald.
\newblock {Singular Lorentz-Violating Lagrangians and Associated Finsler Structures}.
\newblock {\em Phys. Rev. D}, 92(8):085031, 2015.

\bibitem{Sch2}
M.~Schreck.
\newblock {Classical Lagrangians and Finsler structures for the nonminimal fermion sector of the Standard-Model Extension}.
\newblock {\em Phys. Rev. D}, 93(10):105017, 2016.

\bibitem{Ghil1}
D.~M. Ghilencea.
\newblock {Palatini quadratic gravity: spontaneous breaking of gauged scale symmetry and inflation}.
\newblock {\em Eur. Phys. J. C}, 80(12):1147, 4 2020.

\bibitem{Ghil2}
D.~M. Ghilencea.
\newblock {Gauging scale symmetry and inflation: Weyl versus Palatini gravity}.
\newblock {\em Eur. Phys. J. C}, 81(6):510, 2021.

\bibitem{Paulo2}
Adria Delhom, J.~R. Nascimento, Gonzalo~J. Olmo, A.~Yu. Petrov, and Paulo.~J. Porf\'\i{}rio.
\newblock {Metric-affine bumblebee gravity: classical aspects}.
\newblock {\em Eur. Phys. J. C}, 81(4):287, 2021.

\bibitem{Paulo3}
Adri\`a Delhom, J.~R. Nascimento, Gonzalo~J. Olmo, A.~Yu. Petrov, and Paulo.~J. Porf\'\i{}rio.
\newblock {Radiative corrections in metric-affine bumblebee model}.
\newblock {\em Phys. Lett. B}, 826:136932, 2022.

\bibitem{Paulo4}
Adri\`a Delhom, T.~Mariz, J.~R. Nascimento, Gonzalo~J. Olmo, A.~Yu. Petrov, and Paulo~J. Porf\'\i{}rio.
\newblock {Spontaneous Lorentz symmetry breaking and one-loop effective action in the metric-affine bumblebee gravity}.
\newblock {\em JCAP}, 07(07):018, 2022.

\bibitem{Filho:2022yrk}
A.~A. Araújo~Filho, J.~R. Nascimento, A.~Yu. Petrov, and P.~J. Porf\'\i{}rio.
\newblock {Vacuum solution within a metric-affine bumblebee gravity}.
\newblock {\em Phys. Rev. D}, 108(8):085010, 2023.

\bibitem{araujo2024exploring}
A.~A. Araújo~Filho, J.~R. Nascimento, A.~Yu. Petrov, and P.~J. Porf\'\i{}rio.
\newblock An exact stationary axisymmetric vacuum solution within a metric–affine bumblebee gravity.
\newblock {\em Journal of Cosmology and Astroparticle Physics}, 2024(07):004, 2024.

\bibitem{Lambiase:2023zeo}
Gaetano Lambiase, Leonardo Mastrototaro, Reggie~C. Pantig, and Ali Ovgun.
\newblock {Probing Schwarzschild-like black holes in metric-affine bumblebee gravity with accretion disk, deflection angle, greybody bounds, and neutrino propagation}.
\newblock {\em JCAP}, 12:026, 2023.

\bibitem{Jha:2023vhn}
Sohan~Kumar Jha and Anisur Rahaman.
\newblock Study of quasinormal modes, greybody bounds, and sparsity of hawking radiation within the metric-affine bumblebee gravity framework.
\newblock {\em arXiv preprint arXiv:2310.06492}, 2023.

\bibitem{araujo2024gravitational}
A.~A Ara{\'u}jo~Filho, H~Hassanabadi, N~Heidari, J~Kr{\'\i}z, and S~Zare.
\newblock Gravitational traces of bumblebee gravity in metric--affine formalism.
\newblock {\em Classical and Quantum Gravity}, 41(5):055003, 2024.

\bibitem{nascimento2024gravitational}
A.~A Araújo~Filho, J.~R Nascimento, A.~Yu Petrov, and P.~J Porf{\'\i}rio.
\newblock Gravitational lensing by a lorentz-violating black hole.
\newblock {\em arXiv preprint arXiv:2404.04176}, 2024.

\bibitem{lambiase2023probing}
Gaetano Lambiase, Leonardo Mastrototaro, Reggie~C Pantig, and Ali {\"O}vg{\"u}n.
\newblock Probing schwarzschild-like black holes in metric-affine bumblebee gravity with accretion disk, deflection angle, greybody bounds, and neutrino propagation.
\newblock {\em Journal of Cosmology and Astroparticle Physics}, 2023(12):026, 2023.

\bibitem{macedo2015scattering}
Caio~FB Macedo, Ednilton~S de~Oliveira, and Lu{\'\i}s~CB Crispino.
\newblock Scattering by regular black holes: planar massless scalar waves impinging upon a bardeen black hole.
\newblock {\em Physical Review D}, 92(2):024012, 2015.

\bibitem{macedo2016absorption}
Caio~FB Macedo, Luiz~CS Leite, and Lu{\'\i}s~CB Crispino.
\newblock Absorption by dirty black holes: Null geodesics and scalar waves.
\newblock {\em Physical Review D}, 93(2):024027, 2016.

\bibitem{macedo2013absorption}
Caio~FB Macedo, Luiz~CS Leite, Ednilton~S Oliveira, Sam~R Dolan, and Luis~CB Crispino.
\newblock Absorption of planar massless scalar waves by kerr black holes.
\newblock {\em Physical Review D}, 88(6):064033, 2013.

\bibitem{dolan2009scattering}
Sam~R Dolan, Ednilton~S Oliveira, and Lu{\'\i}s~CB Crispino.
\newblock Scattering of sound waves by a canonical acoustic hole.
\newblock {\em Physical Review D}, 79(6):064014, 2009.

\bibitem{crispino2009scattering}
Lu{\'\i}s~CB Crispino, Sam~R Dolan, and Ednilton~S Oliveira.
\newblock Scattering of massless scalar waves by reissner-nordstr{\"o}m black holes.
\newblock {\em Physical Review D}, 79(6):064022, 2009.

\bibitem{crispino2013greybody}
Luis~CB Crispino, Atsushi Higuchi, Ednilton~S Oliveira, and Jorge~V Rocha.
\newblock Greybody factors for nonminimally coupled scalar fields in schwarzschild--de sitter spacetime.
\newblock {\em Physical Review D}, 87(10):104034, 2013.

\bibitem{visser1999some}
Matt Visser.
\newblock Some general bounds for one-dimensional scattering.
\newblock {\em Physical Review A}, 59(1):427, 1999.

\bibitem{boonserm2008bounding}
Petarpa Boonserm and Matt Visser.
\newblock Bounding the bogoliubov coefficients.
\newblock {\em Annals of Physics}, 323(11):2779--2798, 2008.

\bibitem{sakalli2022topical}
{\.I}zzet Sakalli and Sara Kanzi.
\newblock Topical review: greybody factors and quasinormal modes for black holes invarious theories-fingerprints of invisibles.
\newblock {\em Turkish Journal of Physics}, 46(2):51--103, 2022.

\bibitem{leaver1985analytic}
Edward~W Leaver.
\newblock An analytic representation for the quasi-normal modes of kerr black holes.
\newblock {\em Proceedings of the Royal Society of London. A. Mathematical and Physical Sciences}, 402(1823):285--298, 1985.

\bibitem{Pani:2013pma}
Paolo Pani.
\newblock {Advanced Methods in Black-Hole Perturbation Theory}.
\newblock {\em Int. J. Mod. Phys. A}, 28:1340018, 2013.

\bibitem{cardoso2009geodesic}
Vitor Cardoso, Alex~S Miranda, Emanuele Berti, Helvi Witek, and Vilson~T Zanchin.
\newblock Geodesic stability, lyapunov exponents, and quasinormal modes.
\newblock {\em Physical Review D}, 79(6):064016, 2009.

\bibitem{stefanov2010connection}
Ivan~Zh Stefanov, Stoytcho~S Yazadjiev, and Galin~G Gyulchev.
\newblock Connection between black-hole quasinormal modes and lensing in the strong deflection limit.
\newblock {\em Physical review letters}, 104(25):251103, 2010.

\bibitem{wei2011relationship}
Shao-Wen Wei, Yu-Xiao Liu, and Heng Guo.
\newblock Relationship between high-energy absorption cross section and strong gravitational lensing for black hole.
\newblock {\em Physical Review D}, 84(4):041501, 2011.

\bibitem{wei2014establishing}
Shao-Wen Wei and Yu-Xiao Liu.
\newblock Establishing a universal relation between gravitational waves and black hole lensing.
\newblock {\em Physical Review D}, 89(4):047502, 2014.

\bibitem{jusufi2020connection}
Kimet Jusufi.
\newblock Connection between the shadow radius and quasinormal modes in rotating spacetimes.
\newblock {\em Physical Review D}, 101(12):124063, 2020.

\bibitem{liu2020shadow}
Cheng Liu, Tao Zhu, Qiang Wu, Kimet Jusufi, Mubasher Jamil, Mustapha Azreg-A{\"\i}nou, and Anzhong Wang.
\newblock Shadow and quasinormal modes of a rotating loop quantum black hole.
\newblock {\em Physical Review D}, 101(8):084001, 2020.

\bibitem{jusufi2020quasinormal}
Kimet Jusufi.
\newblock Quasinormal modes of black holes surrounded by dark matter and their connection with the shadow radius.
\newblock {\em Physical Review D}, 101(8):084055, 2020.

\bibitem{cuadros2020analytical}
B~Cuadros-Melgar, RDB Fontana, and Jeferson de~Oliveira.
\newblock Analytical correspondence between shadow radius and black hole quasinormal frequencies.
\newblock {\em Physics Letters B}, 811:135966, 2020.

\bibitem{ma2023shadow}
Shi-Jie Ma, Tian-Chi Ma, Jian-Bo Deng, and Xian-Ru Hu.
\newblock Shadow of schwarzschild black hole in the cold dark matter halo.
\newblock {\em Modern Physics Letters A}, 38(24n25):2350104, 2023.

\bibitem{perlick2022calculating}
Volker Perlick and Oleg~Yu Tsupko.
\newblock Calculating black hole shadows: Review of analytical studies.
\newblock {\em Physics Reports}, 947:1--39, 2022.

\bibitem{decanini2011universality}
Yves D{\'e}canini, Gilles Esposito-Farese, and Antoine Folacci.
\newblock Universality of high-energy absorption cross sections for black holes.
\newblock {\em Physical Review D}, 83(4):044032, 2011.

\bibitem{papnoi2022rotating}
Uma Papnoi and Farruh Atamurotov.
\newblock Rotating charged black hole in 4d einstein--gauss--bonnet gravity: Photon motion and its shadow.
\newblock {\em Physics of the Dark Universe}, 35:100916, 2022.

\end{thebibliography}
\bibliographystyle{unsrt}

\end{document}